\newcommand{\be}[0]{\begin{equation}}
\newcommand{\ee}[0]{\end{equation}}
\newcommand{\ba}[0]{\begin{eqnarray}}
\newcommand{\ea}[0]{\end{eqnarray}}

\documentstyle[12pt,epsf]{article}
\input epsf

\begin{document}
\Large
\hfill\vbox{\hbox{IPPP/02/34}\hbox{DCPT/02/68}
\hbox{September 2002}}
\nopagebreak
\vspace{0.75cm}
\begin{center}
\LARGE
{\bf Direct extraction of QCD  ${\Lambda}_{\overline{MS}}$ from moments of structure functions in neutrino-nucleon scattering, using the
CORGI approach}

\vspace{0.6cm}
\Large
C.J. Maxwell$^{1)}$ and  A. Mirjalili$^{2)}$
\vspace{0.4cm}
\large
\begin{em}

Centre for Particle Theory, University of Durham\\
South Road, Durham, DH1 3LE, England
\end{em}
\vspace{1cm}
{\tt {\hspace{10cm}}$^1$) C.J.Maxwell@durham.ac.uk\\
$^2$) Abolfazl.Mirjalili@durham.ac.uk}
\vspace{1cm}
\end{center}
\normalsize
\vspace{0.45cm}
\centerline{\bf Abstract}
\vspace{0.3cm}
We use recently calculated next-to-next-to-leading (NNLO)
anomalous dimension coefficients for the $n=1,3,5,\ldots,13$
moments of the $x{F}_{3}$ structure function in ${\nu}N$ scattering,
together with the corresponding three-loop Wilson coefficients, to
obtain improved QCD predictions for the moments. The Complete
Renormalization Group Improvement (CORGI) approach is used, in
which all dependence on renormalization or factorization scales
is avoided by a complete resummation of ultraviolet logarithms.
The Bernstein Polynomial method is used to compare these QCD
predictions to the $x{F}_{3}$ data of the CCFR collaboration, and
direct fits for ${\Lambda}_{\overline{MS}}^{(5)}$ , with ${N_f}=5$
effective quark flavours , over the range $20<{Q}^{2}<125.9$ ${\rm{Gev}}^{2}$,
were performed. We obtain ${\Lambda}_{\overline{MS}}^{(5)}={202}^{+54}_{-45}$
MeV, corresponding to the three-loop running coupling
${\alpha}_{s}({M}_{Z})=
{0.1174}^{+0.0043}_{-0.0043}$. Including
target mass corrections as well we obtain ${\Lambda}_{\overline{MS}}^{(5)}=
{228}^{+35}_{-36}$ MeV, corresponding to ${\alpha}_{s}({M}_{Z})
={0.1196}^{+0.0027}
_{-0.0031}$.

\
\newpage
\section*{1 Introduction}
The recent measurements of the CCFR collaboration provide the most
precise determination of the deep inelastic scattering (DIS) structure
functions of neutrinos and anti-neutrinos on nucleons \cite{r1}.
In this paper we wish to compare CCFR measurements of the structure function
$x{F}_{3}(x,{Q}^{2})$, with QCD predictions for its moments in order to
determine ${\Lambda}_{\overline{MS}}$ \cite{r2,r3,r3a,r3b}.
We intend to follow essentially the
same method of analysis employed in Refs.\cite{r2,r3}, in which one constructs
averages of the measured structure function with respect to suitably
chosen Bernstein polynomials. The polynomials are chosen so that the
range of $x$ for which the experimental $x{F}_{3}$ is not determined
makes only a small contribution to the averages. These experimental
Bernstein averages are then fitted to the QCD predictions for the corresponding
linear combinations of
 moments. The analysis uses the available
next-next-to-leading order (NNLO) anomalous dimension and coefficient
function results for the odd moments with $n=1,3,5,{\ldots},13$ \cite{r4}.
Refs.\cite{r2,r3} also consider the ${F}_{2}(x,{Q}^{2})$ structure function
in $ep$ DIS at NNLO. The key difference in our analysis is that the
QCD predictions for the moments of $F_3$ are obtained in the
``Complete Renormalization Group Improvement'' (CORGI) approach \cite{r5,r6},
in which all dependence on the renormalization scale ,
$\mu$, and factorization
scale ,
$M$, are eliminated. Customarily in the standard RG-improvement of
QCD predictions for leptoproduction moments one chooses ${\mu}=xQ$ and
$M=yQ$, with $x$ and $y$ undetermined dimensionless constants,
 so the renormalization scale and factorization scale are proportional
to the physical DIS energy scale $Q$. Refs.\cite{r2,r3} make the standard
choice ${\mu}=M=Q$. In the CORGI improvement one instead keeps $\mu$ and $M$
independent of $Q$. One is then forced to resum to all-orders the RG-predictable
ultraviolet (UV)
logarithms of $Q$ , the logarithms of $M$ and $\mu$ contained in
the renormalized coupling constant, and the perturbative coefficients, then
cancel, and one is left with predictions which are independent of ${\mu}$ and
$M$. Crucially in this way one also generates the correct physical $Q$-dependence
of the moments, whereas in standard improvement one omits an infinite subset of
UV logarithms, so that the $Q$-dependence involves the unphysical parameters
$x$ and $y$.
The approach is closely related to the Effective Charge formalism
of Grunberg \cite{r7}. We find that our CORGI fits result in somewhat larger
values of ${\alpha}_{s}(M_Z)$ than those reported in \cite{r3}.

The plan of the paper is to give a brief review of the CORGI approach for
leptoproduction moments in Section 2. Section 3 will contain a short description
of the Bernstein Polynomial averages to be employed in the fits, and Section 4
details the results of the fits to the CORGI predictions for the moments. Section 5
contains a discussion and Conclusions.

\section*{2 Leptoproduction moments in the CORGI approach}
Let us define the Mellin moments for the ${\nu}N$
 structure function
 $
xF_3(x,Q^2)$,
\be
{{\cal{M}}_{n}}(Q^2)=\int_{0}^{1}x^{n-1}F_{3}(x,{Q^2})
\,dx\;.
\ee
Adopting the notation of \cite{r5} we have the factorized form for the ${n}^{\rm{th}}$
moment,
\be
{{\cal{M}}_{n}}(Q^2)=A(n){\left(\frac{ca}{1+ca}\right)}^{d(n)/b}\;\exp({\cal{I}}(a))\;
(1+{r_1}(n){\tilde{a}}+{r_2}(n){\tilde{a}}^{2}+{r_3}(n){\tilde{a}}^{3}+\ldots)\;,
\ee
where the first three terms correspond to the operator matrix element $<{\cal{O}}_{n}(M)>$,
with $M$ the factorization scale, and $a{\equiv}{\alpha}_{s}(M)/{\pi}$ in terms of the
RG-improved
coupling. $A(n)$ is an undetermined scheme-independent overall constant, and
will be one of the parameters varied in the fits.
 ${\cal{I}}(a)$ is a function of the anomalous dimension coefficients ${d}_{i}(n)$,
\be
\frac{M}{<{\cal{O}}_{n}>}\frac{\partial<{\cal{O}}_{n}>}{\partial{M}}={\gamma}_{{\cal{O}}_{n}}=-{d}(n){a}-{d}_{1}(n)
{a}^{2}-{d}_{2}(n){a}^{3}-{\ldots}\;.
\ee
The first anomalous dimension coefficient, $d(n)$,
\cite{r7a}
 is independent of the factorization scheme (FS),
the higher coefficients ${d}_{i}(n)$ $(i{\ge}1)$
define the FS. The coupling $a$ satisfies the beta-function
equation,
\be
M\frac{\partial{a}}{\partial{M}}={\beta}(a)=-b{a^2}(1+ca+{c_2}{a}^{2}+{c_3}{a}^{3}+{\ldots})\;.
\ee
Here $b=(33-2{N_f})/6$ and $c=(153-19{N_f})/12b$ are the first two universal coefficients of the
beta-function. The remaining coefficients are scheme-dependent and determine the renormalization
scheme (RS). The final factor in Eq.(2) is the coefficient function. We use the notation ${\tilde{a}}{\equiv}
{\alpha}_{s}(\mu)/{\pi}$ , where $\mu$ is the renormalization scale. \\

The self-consistency of perturbation theory means that there is a dependence
${r_k}(n)
({\mu},M,{c_2},{\ldots},{c_k}
;{d_1},{d_2},{\ldots},{d_k})$, on the parameters
specifying the FS and RS \cite{r5}. The coefficient ${r_1}$ depends on the factorization
scale $M$, with,
\be
{r_1}(n)(M)=d(n)\left({\ln}\frac{M}{{\tilde{\Lambda}}}-{\ln}\frac{Q}{{\Lambda}_{{\cal{M}}_{n}}}\right)-\frac{{d_1}(n)}{b}\;.
\ee
Here ${\Lambda}_{{\cal{M}}_{n}}$ is an FS and RS-invariant dimensionful constant associated
with the moment, and it is the second UV logarithm in Eq.(5) which determines the physical $Q$-dependence
of ${\cal{M}}_{n}({Q^2})$. ${\Lambda}_{{\cal{M}}_{n}}$ is directly related to ${\Lambda}_{\overline{MS}}$
of modified minimal subtraction by,
\be
\Lambda_{{\cal{M}}_{n}}
=\Lambda_{\overline{MS}}(\frac{2c}{b})^{(\frac{-c}{b})} exp\left(
\frac{d_1(n)}{bd(n)}+
\frac{r_1(n)}{d(n)}\right)\;,
\ee
where ${d_1}(n)$ is the ${\overline{MS}}$ NLO anomalous dimension coefficient, and ${r_1}(n)$ is computed in the
${\overline{MS}}$ scheme with $M=Q$. The ${(2c/b)}^{-c/b}$ factor corresponds to the standard convention for
defining ${\Lambda}_{\overline{MS}}$ \cite{r8}. Using Eq.(5) one can trade $M$ for ${r_1}$ as a parameter
on which $r_k$
depends, similarly $\mu$ can be traded for ${\tilde{r}}_{1}{\equiv}{r_1}(M={\mu})$. The
self-consistency of perturbation theory allows one to obtain the partial derivatives of the $r_k$
coefficients with respect to the FS and RS parameters ${\{}{r_1},{\tilde{r}}_{1},{c_2},{\ldots},{c_k};
{d_1},{d_2},{\ldots},{d_k}{\}}$. On integrating these we obtain expressions for the $r_k$ as multinomials
in the FS and RS parameters, with scheme-independent constants of integration, $X_k$. Thus at NNLO we have,
\be
{r_2}(n)= \left(\frac{1}{2}-\frac{b}{2d(n)}\right)
{r}_{1}^{2}(n)+\frac{b}{d}{r_1}(n){\tilde{r}}_{1}(n)+
\frac{c{d_1}(n)}{2b}-\frac{{d_2}(n)}{2b}-\frac{d(n){c_2}}{2b}+{X_2}(n)\;.
\ee
The constant of integration ${X_k}(n)$ can only be determined given a ${\rm{N}}^{k}$LO perturbative
calculation, whereas the remaining terms are ``RG-predictable'' and can be obtained from lower orders.
The so-called Complete Renormalization Group Improvement (CORGI) result at
${\rm{N}}^{k}$LO corresponds to resumming to {\it all-orders} the RG-predictable terms,
i.e. those {\it not} involving ${X}_{i}(n),(i>k)$ \cite{r5}. It is easy to show that this is equivalent to
working with standard RG-improvement in an FS and RS in which all the parameters are zero.
In this scheme one has \cite{r5} $a={\tilde{a}}={a}_{0}$, where the CORGI coupling $a_0(n)$
can be expressed in terms of the Lambert $W$ function \cite{r9,r10,r10a}, defined implicitly by
$W(z){\exp}(W(z))=z$,
\ba
{a_0}(n)
&=&-\frac{1}{c[1+{W}_{-1}
({z}_{n}(Q))]}
\nonumber \\
{z}_{n}(Q)&{\equiv}&-\frac{1}{e}{\left(\frac{Q}{{\Lambda}_{{\cal{M}}_{n}}
}\right)}^{-b/c}\;.
\ea
Here the ``$-1$'' subscript denotes the branch of the Lambert $W$ function required
for asymptotic freedom, the nomenclature being that of Ref.\cite{r10}. The
${\Lambda}_{{\cal{M}}_{n}}$ is the scheme-invariant, related to ${\Lambda}_{\overline{MS}}$
by Eq.(6). One obtains the ${\rm{N}}^{k}$LO CORGI result,
\ba
{\cal{M}}_{n}(Q^2)&=&A(n){\left(\frac{c{a_0}(n)}{1+c{a_0}(n)}\right)}^{d(n)/b}\left(1+{X_2(n)}{a}_{0}^{2}(n)+
{X_3(n)}{a}_{0}^{3}(n)+\right.
\nonumber\\
&&\left.{\ldots}+{X_k(n)}{a}_{0}^{k}(n)\right)\;.
\ea
Substituting the explicit expression for ${a}_{0}$ in terms of ${W}_{-1}$ in Eq.(8)
we obtain the NNLO CORGI result,
\ba
{\cal{M}}_{n}(Q^2)&=&A(n){[-{W}_{-1}({z}_{n}(Q))]}^{-d(n)/b}\left(1+{X_2(n)}{a}_{0}^{2}(n)\right)
\nonumber \\
{z}_{n}(Q)&{\equiv}&-\frac{1}{e}{\left(\frac{Q}{{\Lambda}_{{\cal{M}}_{n}}}\right)}^{-b/c}\;.
\ea
So we see that the moment is directly proportional to a power of the Lambert function ${W}_{-1}$.
The NNLO CORGI invariants ${X_2}(n)$ can be computed from the ${\overline{MS}}$ results
for ${r_1}(n),{r_2}(n),{d_1}(n),{d_2}(n)$ \cite{r4}. The NNLO anomalous dimension coefficient
${d_2}(n)$ is only known for odd moments, $n=1,3,5,{\ldots},13$ \cite{r4}. On rearranging Eq.(7)
one has,
\be
{X_2(n)}=\left((\frac{-1}{2}-\frac{b}{2d(n)}){r}_{1}^{2}(n)
-\frac{c{d_1(n)}}{2b}+\frac{d_2(n)}{2b}
+\frac{d(n){c_2}}{2b}+{r_2(n)}\right)\;.
\ee
We tabulate the resulting ${X_2}(n)$ CORGI invariants for $n=3,5,{\ldots},13$, and
the ratio ${\Lambda}_{{\cal{M}}_{n}}/{\Lambda}_{\overline{MS}}$ determined from Eq.(6),
in Table 1. We assume ${N_f}=5$ active quark flavours.
\newpage
\begin{center}{\bf Table~1:} The numerical values of the ratio $\frac{\Lambda_{{\cal{M}}_{n}}}{
\Lambda_{\overline{MS}}}$ and the CORGI invariants
${X_2}(n)$, for the odd moments $n=3,5,\ldots,13$ of $x{F_3}$.
\end{center}
$$\begin{array}{|r|rr|r|}\hline n&\frac{{\Lambda}_{{\cal{M}}_{n}}}
{\Lambda_{\overline{MS}}}&&X_2(n)
\\\hline
3&2.268568660&&-.9283306650\\
5&2.999798808&&1.750455480\\
7&3.489368710&&3.858776378\\
9&3.870927376&&5.663749685\\
11&4.188431945&&7.264781157\\
13&4.462684796&&8.714133383\\
\hline\end{array}$$

\section*{3 Bernstein Averages of moments}
In  phenomenological investigations of structure functions, for a given value of $Q^2$, only  a limited number of experimental points,
 covering a partial range of values of $x$, are available. Therefore,
one cannot directly determine the moments.
 A method devised to deal with this situation
 is to take  averages of the structure function weighted by suitable polynomials.
 We can compare theoretical predictions with experimental results for the Bernstein averages, which are defined by \cite{r2}
\be
F_{nk}(Q^2){\equiv}\int_{0}^{1}dxp_{nk}(x)F_3(x,Q^2)
\ee
where $p_{nk}(x)$ are modified Bernstein polynomials,
\be
p_{nk}(x)=2\frac{\Gamma(n+\frac{3}{2})}{{\Gamma(k+\frac{1}{2})}{\Gamma(n-k+1)}}x^{2k}(1-x^2)^{(n-k)}\;,
\ee
and are normalized to unity, $\int_{0}^{1}dxp_{nk}(x)=1$. Therefore the integral (12) represents an average of the function $F_{3}(x)$ in the region
${\bar{x}}_{nk}-\frac{1}{2}\Delta{x}_{nk}{\leq}x{\leq}{\bar{x}}_{nk}-\frac{1}{2}\Delta{x}_{nk}$ where ${\bar{x}}_{nk}$
is the average of $x$ which is very near to the
maximum of $p_{nk}(x)$,
and $\Delta{x}_{nk}$ is the spread of ${\bar{x}}_{nk}$.
 The key point is that
  values of $F_3$ outside this interval contribute little to the integral (12), as $p_{nk}(x)$ decreases to zero very quickly. So, by suitably choosing
   $n$, $k$, we manage to adjust the region where the average is peaked to that in which we have experimental data \cite{r2}.
   Using the binomial expansion in Eq.(13), it follows that the averages of $F_3$ with $p_{nk}(x)$ as weight functions, can be obtained in terms of odd moments,
\be
F_{nk}=2\frac{{(n-k)!}{\Gamma(n+\frac{3}{2})}}{\Gamma(k+\frac{1}{2})\Gamma(n-k+1)}\sum_{l=0}^{n-k}\frac{(-1)^l}{l!(n-k-l)!}\int_{0}^{1}dxx^{(2(k+l)+1)-1)}{F_3}\;,
\ee
using Eq.(1) then,
\be
F_{nk}=2\frac{{(n-k)!}{\Gamma(n+\frac{3}{2})}}{\Gamma(k+\frac{1}{2})\Gamma(n-k+1)}\sum_{l=0}^{n-k}\frac{(-1)^l}{l!(n-k-l)!}{{\cal{M}}_{2(k+l)+1}}\;.
\ee
For the NNLO QCD fits to be performed we are restricted to considering odd moments of $x{F_3}$ for which
the NNLO anomalous dimension coefficient ${d_2}(n)$ has been computed \cite{r4},
 $n=3,5,{\ldots},13$. We can only include
a Bernstein average, $F_{nk}$,
if we have experimental points covering the whole range
 [${\bar{x}}_{nk}-\frac{1}{2}\Delta{x}_{nk}, {\bar{x}}_{nk}-\frac{1}{2}\Delta{x}_{nk}$] \cite{r2},
  this means that we can use only the 10 averages ${F_{21}}^{\hspace{-.225cm}{(exp)}}(Q^2)$, ${F_{31}}^{\hspace{-.225cm}{(exp)}}(Q^2)$,
   ${F_{32}}^{\hspace{-.225cm}{(exp)}}(Q^2)$, ${F_{41}}^{\hspace{-.225cm}{(exp)}}(Q^2)$, ${F_{42}}^{\hspace{-.225cm}{(exp)}}(Q^2)$,
    ${F_{51}}^{\hspace{-.225cm}{(exp)}}(Q^2)$, ${F_{52}}^{\hspace{-.225cm}{(exp)}}(Q^2)$, ${F_{61}}^{\hspace{-.225cm}{(exp)}}(Q^2)$,
${F_{62}}^{\hspace{-.225cm}{(exp)}}(Q^2)$, and ${F_{63}}^{\hspace{-.225cm}{(exp)}}(Q^2)$ \cite{r3}.
To obtain these experimental averages from the CCFR data for $x{F_3}$ \cite{r1}, we fit $x{F_3}(x,{Q^2})$ for
each bin in ${Q}^{2}$ separately, to the convenient phenomenological expression,
\be
{xF_{3}}^{\hspace{-.12cm}{(phen)}}={\cal{A}}x^{\cal{B}}(1-x)^{\cal{C}}\;,
\ee
this form ensures zero values for ${xF_{3}}$ at $x=0$, and $x=1$. A theoretical justification of Eq.(16) may be found in Ref.\cite{r11}.
Using Eq.(16) with the fitted values of ${\cal{A}},{\cal{B}},{\cal{C}}$, one can then compute
${F}_{nk}^{(exp)}({Q}^{2})$ using Eq.(12), in terms of Gamma functions. The resulting experimental
Bernstein averages are plotted in Figure 1. The errors in the ${F}_{nk}^{(exp)}(Q^2)$ correspond to
allowing the CCFR data for $x{F}_{3}$ to vary within the experimental error bars, including experimental
systematic errors \cite{r1}.
 We have only included data for ${Q}^{2}{\ge}20{\rm{GeV}}^{2}$, and our QCD
fits will assume ${N_f}=5$ active quark flavours. This has the merit of simplifying the analysis by avoiding
evolution through flavour thresholds, whilst only reducing the number of fitted ${F}_{nk}^{(exp)}$ points
by eight.
\begin{figure}
\unitlength1cm \hfil
\put(1.5,0){\epsfxsize=10cm \epsffile{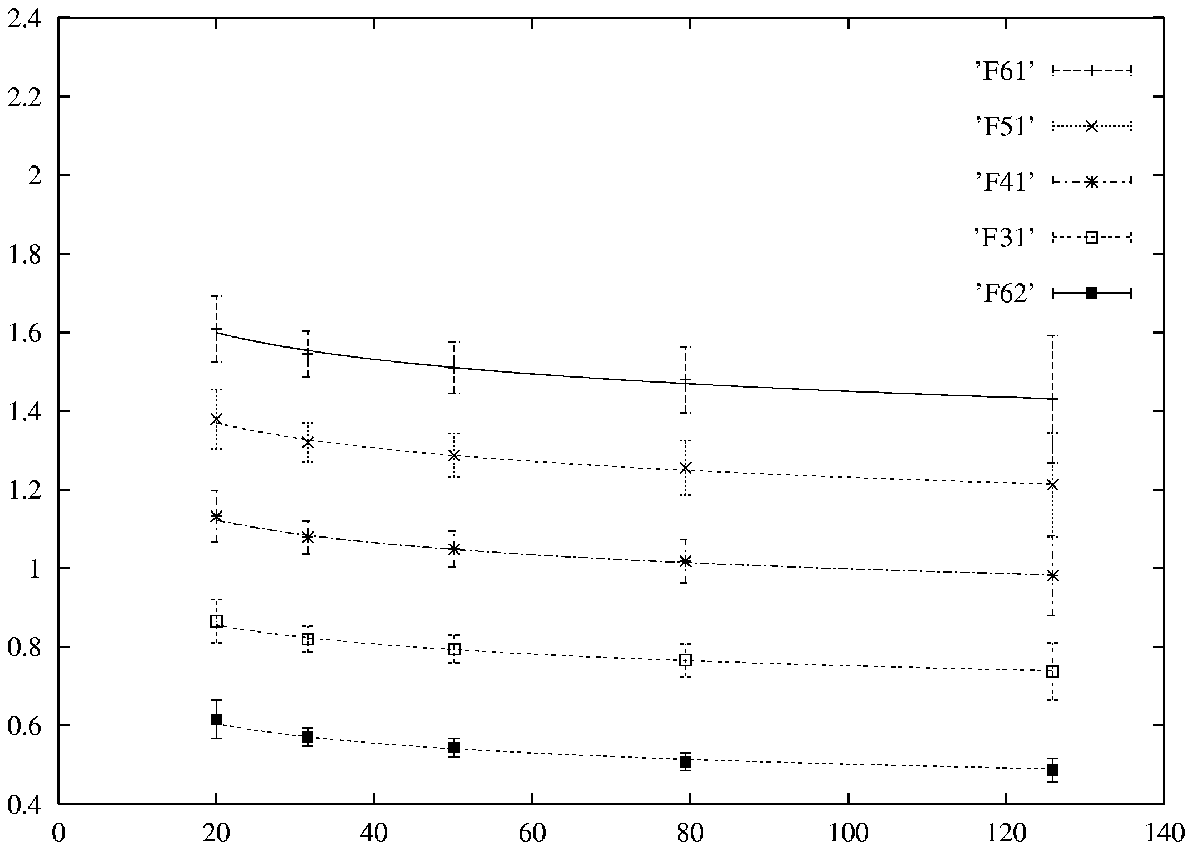}}
\put(1.5,-8){\epsfxsize=10cm \epsffile{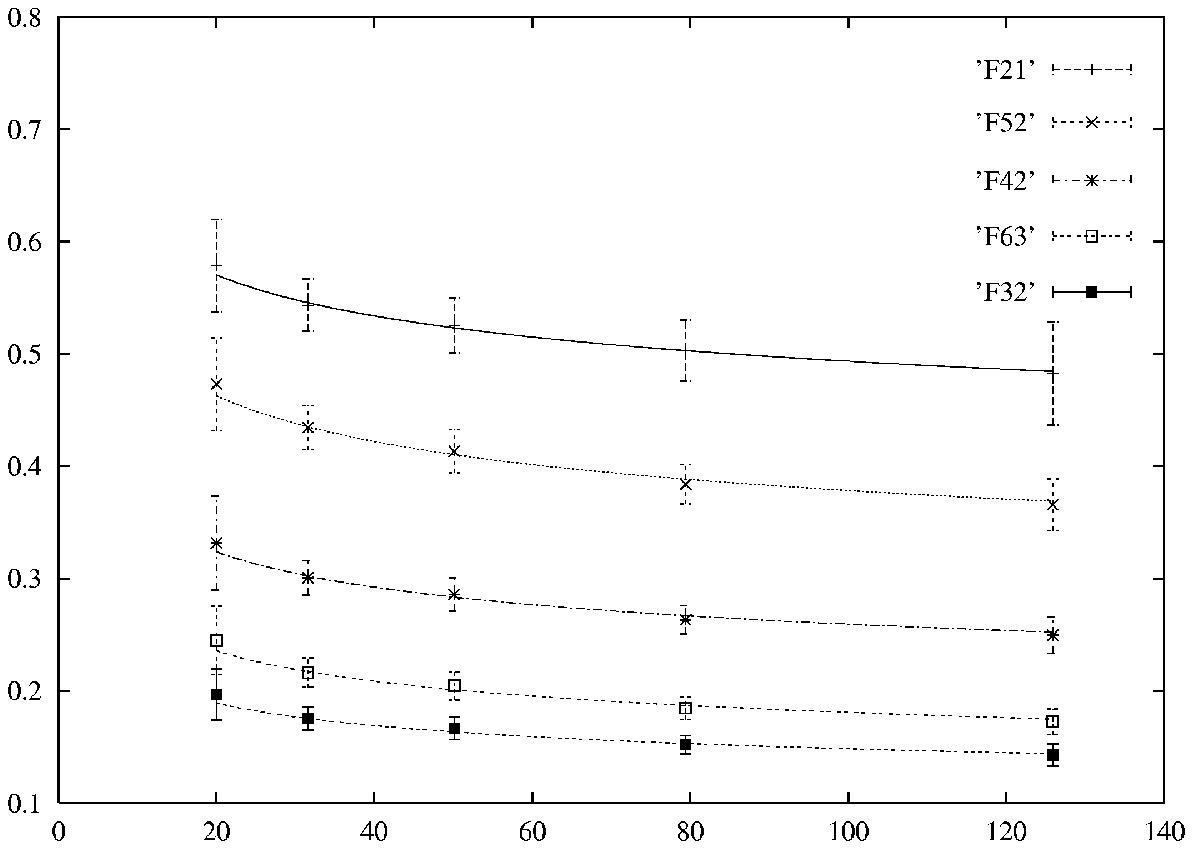}}
\put(1.1,6.6){$F_{nk}$}
\put(6,-.5){$Q^2(Gev^2)$}
\put(1.1,-1.4){$F_{nk}$}
\put(6,-8.5){$Q^2(GeV^2)$}
\caption{Fit to $xF_3$ using Bernstein averages}
\end{figure}
\section*{4 NNLO QCD fits to Bernstein averages for $F_3$}
Using Eq.(15) the ten Bernstein averages ${F}_{nk}({Q}^{2})$ can be written in
terms of odd moments ${\cal{M}}_{n}({Q}^{2})$,
\ba
&&{F_{21}}(Q^2)=7.5\left({\cal{M}}_{3}(Q^2)-{\cal{M}}_{5}(Q^2)\right)
\nonumber\\
&&{F_{31}}(Q^2)=13.125\left({\cal{M}}_{3}(Q^2)-2{\cal{M}}_{5}(Q^2)+{\cal{M}}_{7}(Q^2)\right)
\nonumber\\
&&{F_{32}}(Q^2)=17.5\left({\cal{M}}_{5}(Q^2)-{\cal{M}}_{7}(Q^2)\right)
\nonumber\\
&&{F_{41}}(Q^2)=19.687\left({\cal{M}}_{3}(Q^2)-3{\cal{M}}_{5}(Q^2)+3{\cal{M}}_{7}(Q^2)-{\cal{M}}_{9}(Q^2)\right)
\nonumber\\
&&{F_{42}}(Q^2)=39.375\left({\cal{M}}_{5}(Q^2)-2{\cal{M}}_{7}(Q^2)+{\cal{M}}_{9}(Q^2)\right)
\nonumber\\
&&{F_{51}}(Q^2)=27.070\left({\cal{M}}_{3}(Q^2)-4{\cal{M}}_{5}(Q^2)+6{\cal{M}}_{7}(Q^2)-{\cal{M}}_{11}(Q^2)\right)
\nonumber\\
&&{F_{52}}(Q^2)=72.187\left({\cal{M}}_{5}(Q^2)-3{\cal{M}}_{7}(Q^2)+3{\cal{M}}_{9}(Q^2)-4{\cal{M}}_{9}(Q^2)\right)
\nonumber\\
&&{F_{61}}(Q^2)=35.191\left({\cal{M}}_{3}(Q^2)-5{\cal{M}}_{5}(Q^2)+10{\cal{M}}_{7}(Q^2)-{\cal{M}}_{9}(Q^2)\right.
\nonumber\\
&&\left.{\hspace{1.8cm}}+5{\cal{M}}_{11}(Q^2)-{\cal{M}}_{13}(Q^2)\right)
\nonumber\\
&&{F_{62}}(Q^2)=117.30\left({\cal{M}}_{5}(Q^2)-4{\cal{M}}_{7}(Q^2)+6{\cal{M}}_{9}(Q^2)-4{\cal{M}}_{11}(Q^2)\right.
\nonumber\\
&&\left.{\hspace{1.8cm}}+{\cal{M}}_{13}(Q^2)\right)
\nonumber\\
&&{F_{63}}(Q^2)=187.69\left({\cal{M}}_{7}(Q^2)-3{\cal{M}}_{9}(Q^2)+3{\cal{M}}_{11}(Q^2)-{\cal{M}}_{13}(Q^2)\right)
\ea
We shall use the NNLO CORGI result of Eq.(10) for the QCD prediction of ${\cal{M}}_{n}({Q}^{2})$. The basic fit parameters
will be the unknown normalization constants $A(n)$, $n=3,5,7,\ldots,13$, and ${\Lambda}_{\overline{MS}}$ , related
to the CORGI ${\Lambda}_{{\cal{M}}_{n}}$ by Eq.(6), see Table 1.
Thus there are 7 parameters to be
simultaneously fitted to the experimental ${F}_{nk}({Q}^{2})$ averages. Defining a global ${\chi}^{2}$ for all the
experimental data points of Figure 1 , we found an acceptable fit with minimum
 ${\chi}^{2}/{\rm{d.o.f.}}=1.2905/43$. The best fit is indicated by the curves in Figure 1.
  Allowing ${\chi}^{2}$ within $1$ of the minimum to estimate an error gives,
\be
\Lambda_{\overline{MS}}^{(5)}={202}^{+54}_{-45}{\hspace{.1cm}}{\rm{MeV}}\;,
\ee
which corresponds to the three-loop ${\overline{MS}}$ coupling at the $Z$-mass,
\be
{\alpha}_{s}({M_Z})={
0.1174}^{+0.0043}_{-0.0043}\;.
\ee
The minimum ${\chi}^{2}$ values for all 7 fitting parameters are tabulated in Table 2.
To attempt to include target mass corrections (TMC) in the fits we amended the expression for
${\cal{M}}_{n}({Q}^{2})$ to \cite{r12},
\ba
{\cal{M}}_{n}
^{\hspace{-.035cm}{TMC}}(Q^2)&=&{\cal{M}}_{n}
(Q^2)+\frac{n(n+1)}{n+2}{\frac{m_p^2}{Q^2}}{\cal{M}}_{n+2}
(Q^2)+
\nonumber\\
&&\frac{(n+2)(n+1)}{2(n+2)}{\frac{m_p^4}{Q^4}}{\cal{M}}_{n+4}
(Q^2)+O({\frac{m_p^6}{Q^6}})\;.
\ea
This results from the series expansion in powers of $\frac{{m}_{p}^{2}}{{Q}^{2}}$ of the Nachtmann moments \cite{r13}.
Here $m_p$ is the proton mass. The influence of the O($\frac{{m}_{p}^{4}}{{Q}^{4}}$) terms is very small, and we
shall neglect them. Minimizing ${\chi}^{2}$ with the amended expression for the moments then gives a best fit of
comparable quality, with
\be
\Lambda_{\overline{MS}}^{\hspace{-.11cm}{(TMC)}}={228}^{+35}_{-36}{\hspace{.1cm}}{\rm{MeV}}\;.
\ee
Corresponding to the coupling at the $Z$-mass,
\be
{{\alpha}_{s}({M_Z})}=
{0.1196}^{+0.0027}_{-0.0031}\;.
\ee
The best fit values of the 7 fitting parameters including TMC, are tabulated in Table 3.
We can compare this value of ${\alpha}_{s}(M_Z)$ with the corrresponding values
${\alpha}_{s}(M_Z)=0.1153{\pm}0.0041$ obtained in Ref.\cite{r3}, and ${\alpha}_{s}(M_Z)=0.1187{\pm}0.0026$
obtained in Ref.\cite{r3a}.
Finally we can confirm the expectation that at the energy scales ${Q}^{2}{\ge}20{\rm{GeV}}^{2}$
included in our fit, higher-twist (HT) effects should be small. We modified the
expression for ${\cal{M}}_{n}({Q}^{2})$ by an additional term \cite{r2},
\be
{\cal{M}}_{n}^{(HT)}({Q}^{2})=n\left(\frac{\rho{\Lambda}_{\overline{MS}}^{2}}{{Q}^{2}}\right){\cal{M}}_{n}({Q}^{2})\;,
\ee
where $\rho$ is an additional phenomenological parameter which will be fitted to the data. The best 8 fit
parameters are tabulated in Table 4. The inclusion of HT terms shifts the central value of
${\Lambda}_{\overline{MS}}^{(5)}$ by only $2$ Mev (cf. Table 2) , confirming as expected that HT
effects are negligible. Our fitted value of the HT parameter ${\rho}=-0.77{\pm}0.23$ is to be
compared with the value $-0.14{\pm}0.6$ obtained in Ref.\cite{r3}, and $-0.31{\pm}0.80$ obtained
in Ref.\cite{r3a}.

We should stress that the errors in the values of ${\alpha}_{s}(M_Z)$ quoted in Eqs.(19),(22),
reflect the errors in the ${F}_{nk}^{(exp)}$
 values in Fig. 1 , to which the NNLO CORGI
predictions for the moments of Eq.(10) have been fitted. In the CORGI approach all
dependence on the unphysical factorization and renormalization scales is eliminated by the
complete resummation of RG-predictable UV logarithms, the remaining uncertainty in the
QCD prediction then resides in the unknown ${\rm{N}}^{3}$LO CORGI invariant ${X}_{3}(n)$.
It would in principle be straightforward to use Pad{\'e} approximants to estimate the
unknown ${d}_{3}(n)$ anomalous dimension coefficients, as in Ref.\cite{r3a}, and perform
fits with an estimated ${X}_{3}(n)$, but we have not done so in this work.
\begin{center}{\bf Table~2:}Numerical values of fitting
parameters, for the best fit of Figure 1.
 \end{center}
$$\begin{array}{|r|r|}\hline
\Lambda_{\overline{MS}}&{202}^{+54}_{-45}\;{\rm{MeV}} \\
\hline
A(3)&{0.497}^{-0.018}_{+.022}\\
\hline
A(5)&{0.159}^{-0.008}_{+0.008}\\
\hline
A(7)&{0.07}^{-0.0013}_{+0.0012}\\
\hline
A(9)&{0.04}^{+0.0014}_{-0.0013}\\
\hline
A(11)&{0.03}^{+0.0010}_{-0.009}\\
\hline
A(13)&{0.029}^{-0.0020}_{+0.0019}\\
\hline\end{array}$$\\
\begin{center}{\bf Table~3:}Numerical values of fitting
parameters, including TMC.
 \end{center}
$$\begin{array}{|r|r|}\hline
\Lambda_{\overline{MS}}&{228}^{+35}_{-36}\;{\rm{MeV}}\\
\hline
A(3)&{0.481}^{-0.011}_{+0.012}\\
\hline
A(5)&{0.16}^{-0.004}_{+0.004}\\
\hline
A(7)&{0.08}^{-0.0024}_{+0.0021}\\
\hline
A(9)&{0.05}^{-0.0030}_{+0.0027}\\
\hline
A(11)&{0.03}^{-0.0038}_{+0.0033}\\
\hline
A(13)&{0.009}^{-0.0035}_{+0.0029}\\
\hline\end{array}$$\\
\newpage
\begin{center}{\bf Table~4:}Numerical values of fitting
parameters, including HT.
 \end{center}
$$\begin{array}{|r|r|}\hline
\Lambda_{\overline{MS}}&{204}^{+53}_{-46}\;{\rm{MeV}}\\
\hline
A(3)&{0.497}^{-0.017}_{+.023}\\
\hline
A(5)&{0.159}^{-0.008}_{+0.008}\\
\hline
A(7)&{0.07}^{-0.0013}_{+0.0012}\\
\hline
A(9)&{0.04}^{+0.0014}_{-0.0013}\\
\hline
A(11)&{0.03}^{+0.0010}_{-0.009}\\
\hline
A(13)&{0.029}^{-0.0020}_{+0.0029}\\
\hline
\rho&{-0.77}^{+0.23}_{-0.23}\\
\hline\end{array}$$\\
\section*{5 Discussion and Conclusions}
In this paper we have used a similar method of analysis to that of \cite{r3} to fit
QCD predictions for the moments of the $\nu{N}$ DIS structure function $x{F}_{3}$, to suitably constructed
Bernstein polynomial averages of the CCFR experimental data for $x{F}_{3}$ \cite{r1}. The key difference in
our approach has been the use of Complete Renormalization Group Improved (CORGI) \cite{r5,r6}
NNLO QCD predictions, hence avoiding the need to make particular {\it ad hoc} choices of the
dimensionful renormalization scale, $\mu$, and factorization scale, $M$. The most important
motivation for the CORGI approach is that by completely resumming all the UV logarithms one correctly
generates the {\it physical} dependence of the moments ${\cal{M}}_{n}({Q}^{2})$
on the DIS energy scale $Q$. From Eq.(10) this is seen to be,
\be
{\cal{M}}_{n}({Q}^{2}){\approx} A(n){\left[-{W}_{-1}\left(-\frac{1}{e}{\left(\frac{Q}{{\Lambda}_{{\cal{M}}_{n}}}
\right)}^{-b/c}\right)\right]}
^{-d(n)/b}\left(1+O{\left(\frac{1}{b{\ln}(Q/{\Lambda}_{{\cal{M}}_{n}})}\right)}^{2}\right)\;,
\ee
so that the large-$Q$ behaviour is controlled by the Lambert ${W}_{-1}$ function, and the ratio
$Q/{\Lambda}_{{\cal{M}}_{n}}$, with ${\Lambda}_{{\cal{M}}_{n}}$ the FS and RS-invariant defined
in Eq.(5). In contrast with standard RG-improvement and choosing ${\mu}={M}=xQ$, with $x$ a dimensionless
constant, one has instead,
\be
{\cal{M}}_{n}({Q}^{2}){\approx} A(n){\left[-{W}_{-1}\left(-\frac{1}{e}{\left(\frac{xQ}{{\tilde{\Lambda}}_{\overline{MS}}}
\right)}^{-b/c}\right)\right]}
^{-d(n)/b}\left(1+O{\left(\frac{1}{b{\ln}(xQ/{\tilde{\Lambda}}_{\overline{MS}})}\right)}\right)\;,
\ee
which manifestly depends on the unphysical parameter $x$. Here
${\tilde{\Lambda}}_{\overline{MS}}{\equiv}
{(2c/b)}^{-c/b}$
\newline
$\times{\Lambda}_{\overline{MS}}$.
 In Refs.\cite{r2,r3,r3a} the standard choice
$x=1$ is made. \\

It seems clear to us that CORGI QCD predictions should be used in the fits. However, the value
of ${\alpha}_{s}(M_Z)=0.1196{\pm}0.003$ that we obtain, including TMC effects, is consistent with the
result ${\alpha}_{s}(M_Z)=0.1153{\pm}0.0063$ obtained in Ref.\cite{r3} using a similar method of
analysis, but with the standard $x=1$ scale choices. We are in even closer agreement with the
value ${\alpha}_{s}(M_Z)=0.1195{\pm}0.004$ reported in Ref.\cite{r3a}, but these authors use a
very different method of analysis involving the Jacobi Polynomial technique, together with the standard
$x=1$ scale choice.
 This indicates that the incomplete resummation of UV logarithms
implicit in the standard scale choice ${\mu}=M=Q$,
 does not greatly modify the fit. This is underwritten
by the reasonably small NNLO CORGI invariants ${X}_{2}(n)$, and ratios ${\Lambda}_{{\cal{M}}_{n}}/{\Lambda}_
{\overline{MS}}$, appearing in Table 1. From Eq.(6) and Table 1 one can see that the CORGI result
corresponds to using an $n$-dependent scale choice, ranging from $x=0.44$ for
the $n=3$ moment to
$x=0.22$ for the $n=13$ moment. The CORGI approach also differs in that ${c}_{2}=0$ and
${d}_{1}(n)={d}_{2}(n)=0$, rather than the $\overline{MS}$ values.
 It should be noted, however, that use of the standard physical
choice of renormalization scale is in general likely to result in misleading determinations
of ${\Lambda}_{\overline{MS}}$, as has been discussed in Ref.\cite{r6} in the case of
${e}^{+}{e}^{-}$ jet observables.

\section*{Acknowledgements}
We are grateful to Siggi Bethke for pointing out that the quoted values of the three-loop
${\alpha}_{s}(M_Z)$ in an earlier version of this paper were not consistent with the
reported ${\Lambda}_{\overline{MS}}$ values.
A.M. is grateful to the Institute for Particle Physics Phenomenology (IPPP) of Durham University
for their hospitality whilst this research was performed, and thanks K. Javidan for useful
discussions.

\end{document}